\documentclass[aps,prl,twocolumn,showpacs,preprintnumbers,amsmath,amssymb,groupedaddress]{revtex4-1}

\usepackage{soul}

\usepackage{graphicx}
\usepackage{dcolumn}
\usepackage{bm}
\usepackage{color}
\definecolor{Red}  {rgb}{1,0,0}
\definecolor{Green}{rgb}{0,1,0}
\definecolor{Blue} {rgb}{0,0,1}

\newcommand {\bfv}[1] {{\boldsymbol {#1}}}

\newcommand\Rey{\mbox{\textit{Re}}}  

\newcommand\removed[1]{}

\newcommand\SEC[1]{\vskip 1mm {\it #1}.\rule[1mm]{5mm}{0.1mm}}
\newcommand{\gtsim}{\protect\raisebox{-0.5ex}{$\:\stackrel{\textstyle >}{\sim}\:
$}}
\usepackage{ulem}

\def\Ikaku#1{{\color{Blue}{#1}}}

\usepackage{slashbox}

\begin{document}

\title{On the Spiral Roll State in Heat Convection between Non-Rotating Concentric Double Spherical Boundaries}


\author{}
\affiliation{}

\author{Takahiro Ninomiya}
\affiliation{
  Department of Pure and Applied Physics,
  Faculty of Engineering Science, Kansai University,  Osaka, 564-8680, Japan
}
\author{Keito Konno}
\affiliation{
  Department of Pure and Applied Physics,
  Faculty of Engineering Science, Kansai University,  Osaka, 564-8680, Japan
}
\author{Masako Sugihara-Seki}
\affiliation{
  Department of Pure and Applied Physics,
  Faculty of Engineering Science, Kansai University,  Osaka, 564-8680, Japan
}
\author{Tomoaki Itano}
\affiliation{
  Department of Pure and Applied Physics,
  Faculty of Engineering Science, Kansai University,  Osaka, 564-8680, Japan
}


\date{\today}

\begin{abstract}
\ \ \ We study the single-arm spiral roll state in the system of Boussinesq fluid confined between non-rotating double concentric spherical boundaries with an opposing temperature gradient previously reported by Zhang et al.(2002)\cite{Zha02}.
It is found that the state exists even in a fairly thicker gap than that used in this previous study, as an autonomously rotating wave solution with a finite constant angular velocity in non-rotating geometry.
We further find that individual spiral states bifurcate directly from the static state at a number of intersections of the marginal stability curves.
\end{abstract}



\maketitle

\SEC{Introduction}
Recent global seismological observation has established the presence of fluid core at the centre of our planet, which may be capable of explaining several hitherto unsolved phenomena\cite{Fow04}.
Such evidence provides a convincing grounds for the proposal that thermal convection in the earth's core is the primary mechanism inside the apparently solid but pulsating planet, even though direct measurement of convection is still beyond our reach.
While more realistic configurations are thought to be necessary to model the convection in the core and to reproduce the autonomous secular variations of terrestrial geomagnetic field\cite{Kid97,Kim11}, the variety and the beauty of the possible convective patterns that may be found even in simpler geometries has been perceived only by a handful of pioneers in the field of mathematical science.
One of idealised problems which allow us to model the convection is the system given by Boussinesq fluid confined between non-rotating double concentric spherical boundaries with an opposing temperature gradient, which has been extensively investigated over the past few decades following Chandrasekhar's introduction\cite{Cha61}.

Due to the spherical homogeneity inherited in the system, at the transition from the thermal conductive to convective states, the degenerated unstable modes emerge simultaneously, so that either axisymmetric or relatively highly-symmetric steady states invariant under a set of transformations of point groups may bifurcate directly from the static state via some nonlinear interactions\cite{Bus75,Zeb83,Feu11}.
Although it had been so far intuitively believed that such highly-symmetric steady states would dominate either other less-symmetric equilibria or asymmetric chaotic states in the transitional stage, recent numerical studies\cite{Zha02,Li05} reported that a less-symmetric equilibrium state, namely a {\it single-arm spiral roll state}, coexists with other equilibrium states and appears to be stable.
Bearing in mind that symmetry-breaking initial conditions are common in nature, the possibility that less symmetric states are preferred in practice thus merits further investigation.
In what follows, we will describe interesting and unexpected aspects of the spiral roll state, which have hitherto remained obscure.

    \begin{figure}
      \centerline{
        \includegraphics[angle=0,width=0.50\textwidth]{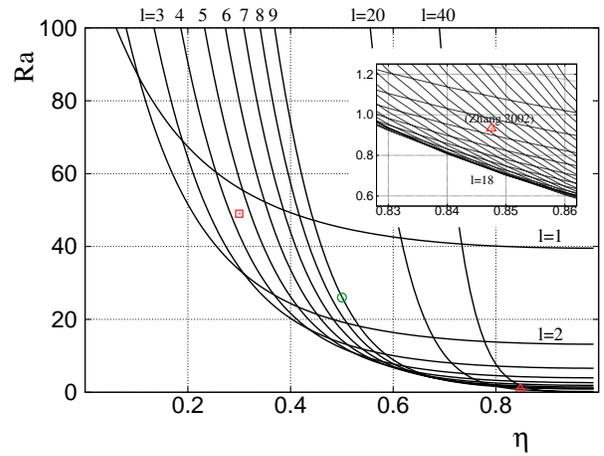}
      }
      \caption{
        Neutral curves of the static state, $Ra=Ra(\eta,l)$.
        For a fixed value of $\eta$, the static state is asymptotically stable against an infinitesimal disturbance composed of $l$-th degree spherical harmonic functions at $Ra<Ra_{\rm cr}(\eta):={\rm min} \{ Ra(\eta,l) ; l=1,2,\cdots \}$.
        In the inset plot, the triangle symbol plotted at $(\eta,Ra)=(0.847,0.933)$ indicates the position at which a relatively stable spiral roll state is obtained from the numerical simulation conducted by Zhang's group \cite{Zha02,Li05} (see also Fig.\ref{fig2}(right)).
        Other symbols correspond to the parameters where the spiral states shown in Fig.\ref{fig2}(left) and Fig.\ref{fig4} are obtained in the present study.
      }
      \label{fig1}
    \end{figure}

    \begin{figure}
      \centerline{
        \includegraphics[angle=0,width=0.25\textwidth]{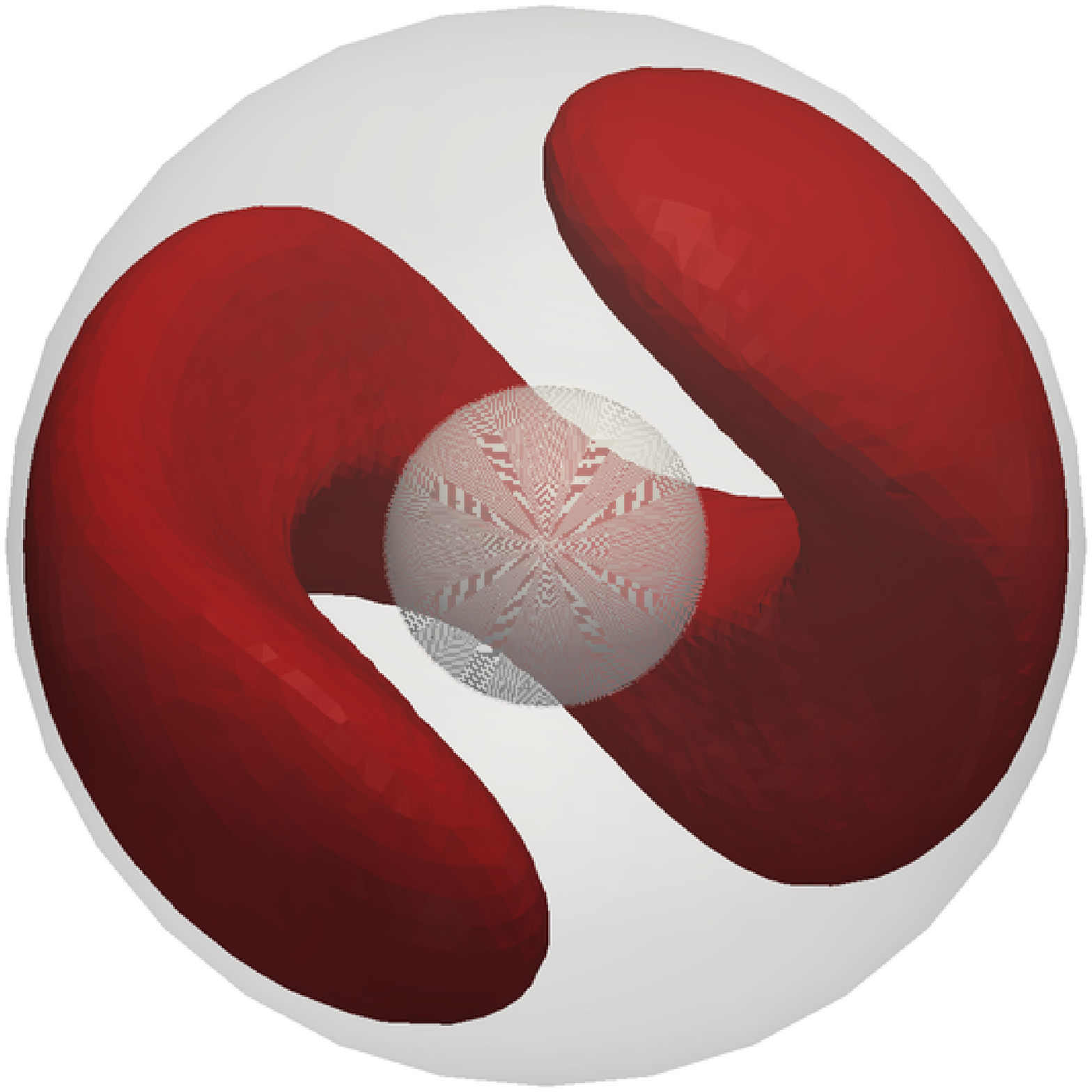}
        \includegraphics[angle=0,width=0.25\textwidth]{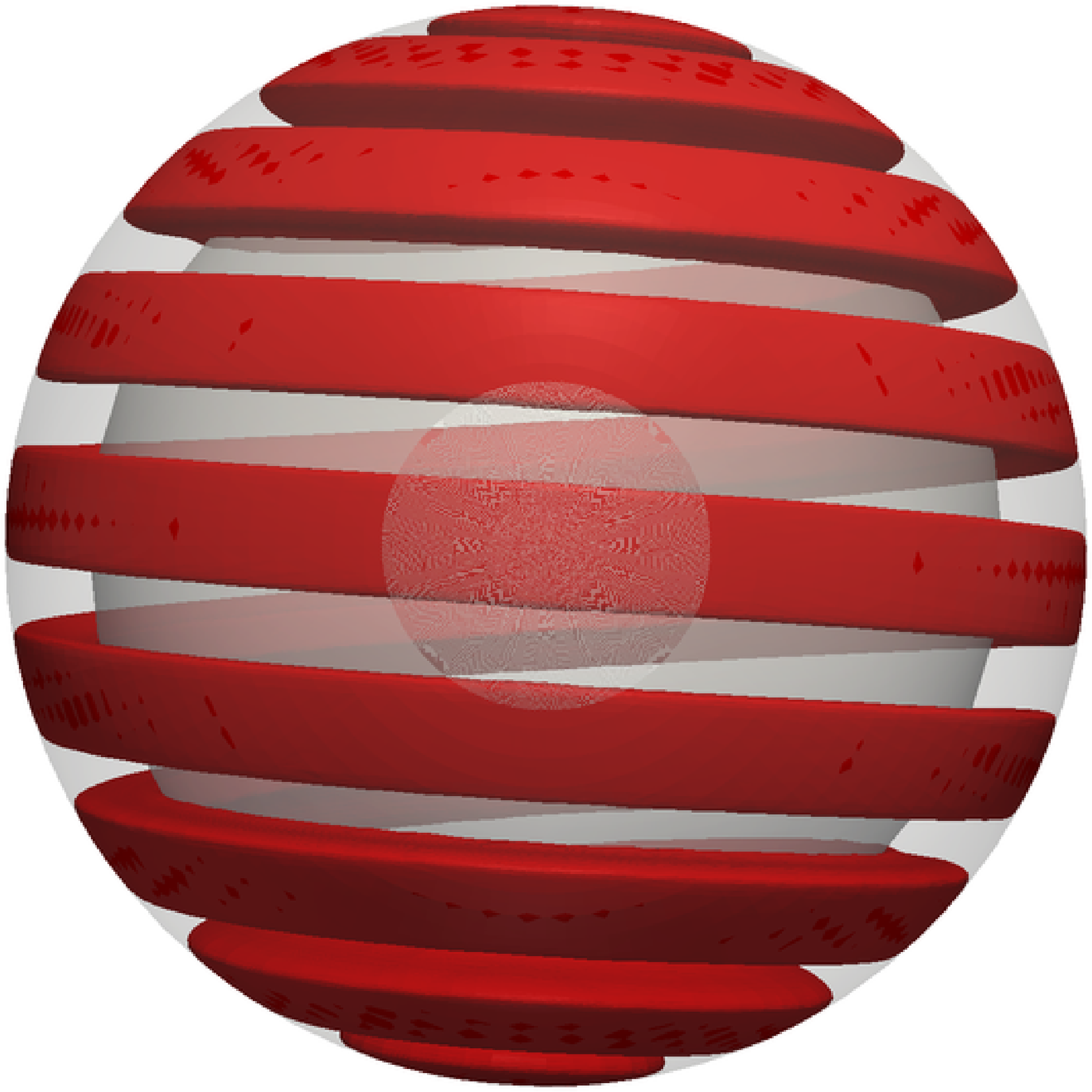}
      }
      \caption{
        The spiral states obtained at $Ra$ slightly larger than the critical Rayleigh number, $Ra_{\rm cr}:={\rm min}_l Ra(\eta,l)$, for different values of $\eta$, are visualised by an isosurface of thermal deviation $\varTheta=0.1$ .
        The corresponding parameter values are indicated as red circles in Fig. \ref{fig1}, (left) $(Ra,\eta)=(50,0.30)$ and (right) $(Ra,\eta)=(0.933,0.847)$. The latter probably corresponds to that obtained in Ref.\cite{Zha02}.
      }
      \label{fig2}
    \end{figure}

\SEC{Formulation}
Firstly following Chandrasekhar\cite{Cha61}, under the assumption that the distributions of mass and internal heat generation inside the outer spherical boundary are uniform, we obtain the following nondimensionalised form of the governing equations of the system,
\begin{eqnarray*}
  \partial_t \bfv{u} + (\bfv{u}\cdot\bfv{\nabla})\bfv{u} &=& - \bfv{\nabla} p + Ra  \varTheta r \bfv{e}_r + \bfv{\nabla}^2 \bfv{u} \ \ , \\
  Pr \bigl( \partial_t \varTheta + \bfv{u}\cdot\bfv{\nabla} \varTheta \bigr) &=& r u_r  + \bfv{\nabla}^2 \varTheta \ \ ,
\end{eqnarray*}
 where $Ra$ and $Pr$ are Rayleigh and Prandtl numbers, and $\bfv{u}$, $p$ and $\varTheta$ denote the flow velocity, pressure and temperature deviation fields, respectively.
While the second term of the right-hand side of the first equation follows the proportionality of the gravity acceleration to the radius, the first term on the right-hand side of the second equation arises due to the quadratic polynomial form of the temperature field in the static state.
Non-slip and isothermal boundary conditions are imposed at the concentric inner and outer spherical boundaries, and the half-width of the gap is taken to be the unity length scale in our non-dimensionalisation.
Thus, the system is determined only by the three control parameters, $Ra$, $Pr$ and the radius ratio of the inner to outer boundaries, $\eta$ ($0<\eta<1$).
The velocity field, $\bfv{u}$, can be decomposed into toroidal $\Psi$ and poloidal $\Phi$ components with regard to the radial direction due to the incompressiblity constraint.
Hereafter we will neglect the former component because of the absence of an energy source term that is required to sustain it \cite{Bus75}.

\SEC{Linear stability of the static state}
The marginal stability (neutral) curves of the static conductive state on the map of $(\eta,Ra)$ are shown in Fig.\ref{fig1}, which is obtained by solving the eigenvalue problem deduced via the linearization from the aforementioned governing equations.
The conductive state loses its stability independently of the value of $Pr$ to the right of the neutral curves shown in the figure.
These curves are distinguished by the degree $l$ of degenerated spherical harmonic functions in which the superimposed infinitesimal perturbation is decomposed.
It should be noted that, in the narrow gap limit, our system is identical to the so called Rayleigh-B\'enard convection in a planar layer between horizontal parallel plates of infinite extent.
As a spherical harmonic function of degree $l$ may represent $l$ eddies in the meridian section from the north to the south poles, the length of a meridian line is approximated by the product of the gap width with the number of the eddies. 
This estimation gives us a reasonable approximation of the most dangerous degree $l_{\rm cr}$ as the function of $\eta$ in the limit $\eta\to 1$, $l_{\rm cr}(\eta) \approx {\rm O}\Bigl(\frac{1+\eta}{1-\eta}\Bigr)$.
As a result, a higher degree spherical harmonic function is associated with the characteristic convective state at higher $\eta$.

\SEC{Existence of spiral roll state in wide gaps}
The triangle symbol plotted at $(\eta,Ra)=(0.847,0.933)$ in the figure indicates the position at which it was confirmed numerically by Zhang's group \cite{Zha02,Li05} that a single-arm spiral roll state is a steady and stable state for a variety of values of $Pr$, the range $10^{-1}<Pr<10^{2}$.
For fixed values of $\eta$, the static state first loses its stability at the critical Rayleigh number $Ra_{\rm cr}=0.729$ for a perturbation of the degree $l=18$, and is stable to perturbations of other degrees.
However, with a slight increase of $Ra$ up to $0.933$, the number of the unstable eigenmodes of the static state drastically increases, as implied by dense curves seen in the inset of the figure.
This would imply that the number of effective degrees of freedom in the system is huge at $\eta \sim 1$ even if the increase of $Ra$ from $Ra_{\rm cr}(\eta)$ is relatively small.
Thus, the spiral roll state would not be expected to be identified as a steady state in long time-interval direct numerical simulations.
This could be why many subspecies of steady spiral states featuring dislocations were reported in the previous investigations\cite{Zha02,Li05}.

On the other hand, according to Ref.\cite{Mat03} suggesting the ubiquitous existence of spiral patterns in a variety of dynamical systems expressed in spherical geometry, $\it e.g.,$ reaction-diffusion system described by the Swift-Hohenberg equation\cite{Swi77} on a sphere surface, a spiral convective state may exist even at lower $\eta$ in our thermal convection system, in general, as a rotating wave solution with a finite angular velocity.
Thus, we have developed a numerical scheme to exactly solve for a spiral roll state in a wider gap than studied previously, restricting our scope to the case $Pr=1$.
With the aid of a numerical library provided by Ref.\cite{Ish13} for the expansion of the spherical harmonic functions, the velocity and temperature deviation fields are expanded into a series of modified Chebyshev polynomials in the radial direction and spherical harmonic functions in the polar and azimuthal directions, respectively truncated at $(N_{\rm radial},N_{\rm spherical})$.
Suppose a set of unknown coefficients to determine the fields is represented by the vector $\bfv{X}$, then the aforementioned governing equations are reduced to ${\it \Omega}^o{\bfv{X}}+{\cal L}(\bfv{X})+{\cal F}(\bfv{X},\bfv{X})=0$, where ${\it \Omega}^o$ is a tensor corresponding to a rotation of the fields determined by an angular velocity, and ${\cal L}$, ${\cal F}$ are linear and nonlinear operators respectively.
By fixing the azimuthal phase of the fields, the above equation can be numerically solved by the Newton-Raphson method, anticipating in general a rotating wave solution to exist at a given set of parameters.

In Fig.\ref{fig2}, equilibrium states solved at $\eta=0.3$ and $0.85$, which correspond to both ends of our scope on $\eta$ in the present survey, are visualised by an isosurface of temperature deviation.
Expanded into a plane spanned by spherical ordinates $(\theta,\phi)$, the state would be apparently consisted by a steady couple of rolls with positive and negative vorticity as if ordinary quasi two-dimensional roll patterns with small dislocations would been observed in the conventional planar Rayleigh-B\'enard system, but the analogy between them is justified only at the narrow gap limit.
It should again be emphasised that the spiral roll state however exists not only in narrow gaps\cite{Zha02} but also at wider gaps, as a rotating wave rather than a steady solution.
Thus, it is natural that the the spiral roll state in a wider gap has a larger angular velocity, while that in a narrow gap is apparently steady.
In addition, it is also plausible that the tips of a roll cause some geometrical instability inducing new dislocations or meandering the roll itself, so that one might intuitively speculate that the spiral roll state could be sustained as an equilibrium state only in the case of narrow gap, where the ratio of tips to the full length of a roll surrounding the sphere is relatively small.

\begin{figure}[t]
      \centerline{
        \includegraphics[angle=0,width=0.50\textwidth]{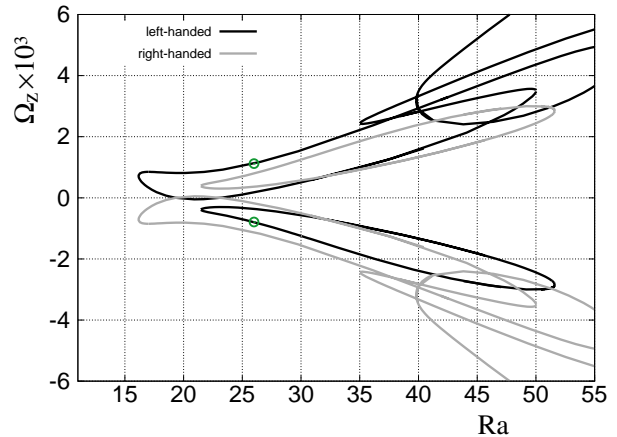}
      }
      \caption{
        The angular velocity of the spiral roll state obtained at $\eta=0.5$ against $Ra$. 
        The black solid curves correspond to the spiral roll pattern screwed in a manner of anti-clockwise rotation along the line connecting both ends of its roll (e.g. states plotted in Fig.\ref{fig2}), while the gray solid curves are those of the corresponding states reflected by the mirroring transformation, where the spiral pattern is screwed in a manner of clockwise rotation.
      }
      \label{fig3}
    \end{figure}

\SEC{Multiplicity of single-arm spiral roll states}
The single-arm spiral roll state satisfies a 2-fold rotational symmetry, ${\cal C}_2$, which means an invariance under a rotation of $\pi$ with respect to a fixed axis, referred as the $z$ axis hereafter\cite{Mat03}.
As described above, the state, in general, evolves with a rotation at a finite constant angular velocity $\Omega_z$ around the $z$ axis.
As illustrated in Fig.\ref{fig3}, the magnitude of ${\Omega}_z$ of the spiral roll state obtained at $\eta=0.5$ is of ${\rm O}(10^{-3})$, which would be sufficiently large to be measurable in numerical simulation, while it tends to become much smaller with increase of $\eta$.
Actually, a spiral state at $(\eta,Ra)=(0.847,0.933)$ reported by Zhang's group is reconfirmed to have a small but non-zero magnitude of $\Omega_z$ of ${\rm O}(10^{-6})$ (Fig.\ref{fig2}(right)).
This could explain why a single-arm spiral roll state was detected as a steady state for relatively large $\eta>0.8$.

It should be also noted that, in the present system, an exact solution can be converted to another exact solution via the mirroring reflection transformation ({\it parity}) due to the spherical homogeneous geometry.
However, the spiral roll state spatially exhibits chirality, which means the presence of an identical exact solution under a non-superposable mirroring transformation, where the sign of the angular velocity is inverted.
In the figure, the black solid curves correspond to the left-handed spiral roll pattern, which thus follows an anti-clockwise rotation along the line connecting both ends of its roll, as presented in Fig.\ref{fig2}, while the gray solid curves are those of the corresponding states reflected by the mirroring transformation, which instead is screwed by a clockwise rotation along a line connecting its ends.
The diversity of the spiral states is attributed partially to the fact that the manner of the screwing (left-handed or right-handed), the type of chirarity, does not determine the sign in the angular velocity.
The bifurcation curves experience a few turning points, so that several different types of spiral roll states with a small variation in their shape may coexist even if we fix the control parameters.
Additionally, it should be also noted that at least two individual branches (black curves) never connects to each other at $\eta=0.5$.

\begin{figure}[t]
      \centerline{
        \includegraphics[angle=0,width=0.25\textwidth]{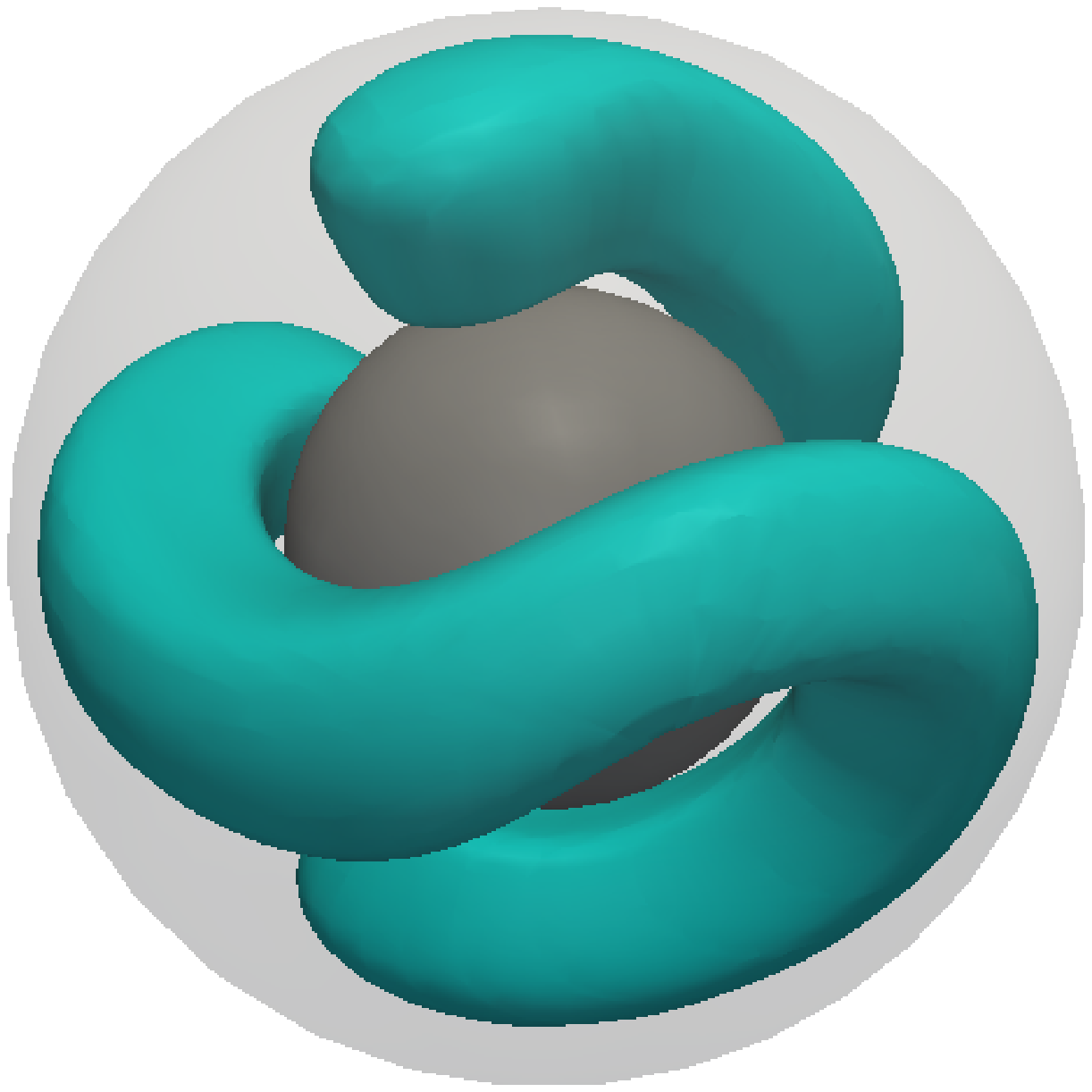}
        \includegraphics[angle=0,width=0.25\textwidth]{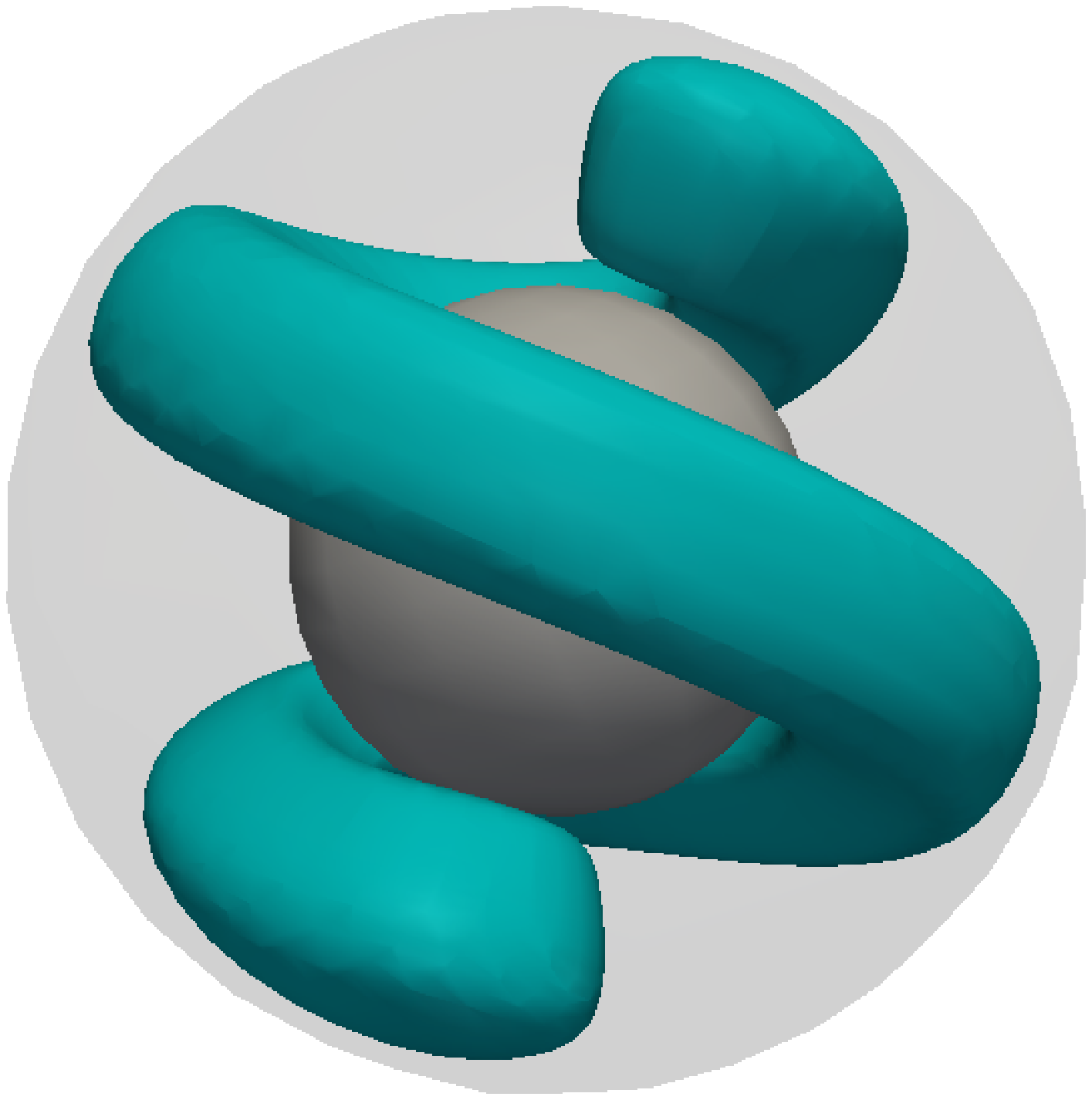}
      }
      \caption{
        The spiral states at $(\eta,Ra)=(0.5,26)$, which correspond to (left) $\Omega_z=1.12\times 10^{-3}$ and (right) $\Omega_z=-7.93\times 10^{-4}$ in Fig.\ref{fig3}, respectively.
      }
      \label{fig4}
    \end{figure}

\SEC{Origin of spiral states}
It was thought that less-symmetric states as a spiral state, in general, do not bifurcate directly from the static state, and that empirically a stable spiral state can be obtained for $Ra > 1.28 \times Ra_{\rm cr}(\eta)$ (see Ref.\cite{Li05}).
However, in the present study, we confirmed that the lowest value of $Ra$ at which an exact spiral roll state exists is not always far from the neutral curves, but instead the spiral state may bifurcate directly from the static state at a point $(Ra,\eta)$ satisfying $Ra(\eta,l)=Ra(\eta,l+1)$ for $l \ge 3$.
The point, which is hereafter referred to as an {\it adjacent node}, is one of intersections of the neutral curves associated with spherical harmonic functions of the adjacent orders.
The obtained numerical evidence provides that a spiral state at an adjacent node is consisted of the superposition of infinitesimal perturbation of spherical harmonics with different but adjacent order, $l$ and $l+1$, which may satisfy ${\cal C}_2$ symmetry and achieve a stable equilibrium saturated by nonlinear interaction.
This can be deduced by accounting for a possible combination of perturbations to the static state at a adjacent node.
Suppose $\bfv{X} = \bfv{X}_0+\epsilon\bfv{X}_1+{\rm O}(\epsilon^2)$ ($|\epsilon| \ll 1$), where the static state $\bfv{X}_0=0$ and $\bfv{X}_1$ is a set of eigenmodes satisfying ${\cal L}(\bfv{X}_1)=0$ at the adjacent node, then the balance up to the second order of the perturbation is written as
\[ \epsilon \Omega^O \bfv{X}_1 + \epsilon^2 {\cal F}(\bfv{X}_1,\bfv{X}_1) = 0 \ .\]
It follows that the magnitude of the angular velocity $\Omega_z$ is of ${\rm O}(\epsilon)$ at the adjacent nodes of neutral curves, and that a branch of the spiral state may emerge directly from the static state via a pitchfork bifurcation.
Moreover, the existence of an infinite number of adjacent nodes would suggest that a number of individual spiral states could have their births at distinct adjacent nodes.
In fact, the individuality of spiral states originated in the different nodes probably reflects that the two branches of same chirality (see Fig.\ref{fig4}) but positive and negative angular velocity do not connect in the diagram of angular velocity as seen in Fig.\ref{fig3}.

\SEC{Conclusion and further works}
\ \ \ Single-arm spiral roll states are obtained in a spherical Rayleigh-B\'enard system with a relatively wider gap than that reported in previous investigations\cite{Zha02,Li05}.
In general, spiral states are described as a rotating wave solution around an axis, with respect to which the solution is invariant under a rotation of $\pi$, but the angular velocity is quite small in the narrow-gap limit.
A number of individual spiral roll states may originate at distinct intersections of the marginal stability curves of the static state.
While the inclusion of the magnetic field and Coriolis force seems to be necessary to reproduce the secular variations of terrestrial geomagnetic field, the presence of the autonomously rotating wave solution in a non-rotating spherical geometry would imply that the mechanism of the variation underlies even a much simpler dynamical system than expected.
This work has been also supported in part by KAKENHI (23760164) and by the Kansai University Special Research Fund 2012. 
T.I. would like to thank Prof. T. Kawahara for his kindly providing private teachings on the subject of nonlinear interaction and to acknowledge Dr D. P. Wall for his conscientious improving the manuscript.

\bibliographystyle{apsrev4-1}
\bibliography{prl05}

\end{document}